\documentclass[aps,prd,reprint,amsmath,amssymb,floatfix]{revtex4-1}
\usepackage[colorlinks]{hyperref}
\usepackage{graphicx}
\usepackage{dcolumn}
\usepackage{booktabs}
\DeclareMathOperator{\re}{Re}
\DeclareMathOperator{\atan2}{atan2}
\usepackage[utf8]{inputenc}
\usepackage[T1]{fontenc}

\newcommand{\Ndot}{\dot{\mathcal{N}}}

\usepackage{acronym}

\newacro{LIGO}[LIGO]{Laser Interferometer Gravitational\nobreakdashes-Wave Observatory}
\newacro{IPN}[IPN]{Interplanetary Network}
\newacro{SNR}[SNR]{signal\nobreakdashes-to\nobreakdashes-noise ratio}
\newacro{SVOM}[\textit{SVOM}]{Space\nobreakdashes-based multi\nobreakdashes-band astronomical Variable Objects Monitor}
\newacro{UFFO-P}[UFFO-P]{Ultra\nobreakdashes-fast Flash Observatory Pathfinder}
\newacro{SGR}[SGR]{soft gamma\nobreakdashes-ray repeater}
\newacro{FAP}[FAP]{false\nobreakdashes-alarm probability}
\newacro{FAR}[FAR]{false\nobreakdashes-alarm rate}
\newacro{CBC}[CBC]{compact binary coalescence}
\newacro{NS}[NS]{neutron star}
\newacro{BH}[BH]{black hole}
\newacro{GRB}[GRB]{gamma\nobreakdashes-ray burst}
\newacro{GW}[GW]{gravitational wave}
\newacro{EM}[EM]{electromagnetic}
\newacro{TOA}[TOA]{time\nobreakdashes-of\nobreakdashes-arrival}
\newacroplural{TOA}[TOAs]{times\nobreakdashes-of\nobreakdashes-arrival}
\newacro{rms}[rms]{root\nobreakdashes-mean\nobreakdashes-square}
\newacro{pdf}[pdf]{probability density function}
\newacro{FOV}[FOV]{field of view}
\newacro{SMT}[SMT]{slewing mirror telescope}
\newacroplural{FOV}{fields of view}
\newacro{CDF}[CDF]{cumulative distribution function}
\newacro{SGR}[SGR]{soft $\gamma$ repeater}
\newacro{H2}[H2]{Hanford 2\nobreakdashes-km detector}
\newacro{L1}[L1]{Livingston 4\nobreakdashes-km detector}
\newacro{S5}[S5]{LIGO's fifth science run}
\newacro{NSNS}[NS\nobreakdashes--NS]{neutron star\nobreakdashes--neutron star}
\newacro{NSBH}[NS\nobreakdashes--BH]{neutron star\nobreakdashes--black hole}
\newacro{BHBH}[BH\nobreakdashes--BH]{black hole\nobreakdashes--black hole}
\newacro{S5VSR1}[S5/VSR1]{\ac{LIGO}'s fifth and Virgo's first science runs}
\newacro{S6VSR23}[S6/VSR2,3]{\ac{LIGO}'s sixth and Virgo's second and third science runs}
\newacro{NSF}[NSF]{National Science Foundation}
\newacro{GBM}[GBM]{$\gamma$\nobreakdashes-ray Burst Monitor}
\newacro{BAT}[BAT]{Burst Alert Telescope}
\def\mnras{\ref@jnl{MNRAS}}             
\def\apjl{\ref@jnl{ApJ}}                

\begin{document}

\title{Outlook for detection of \acs{GW} inspirals by \acs{GRB}-triggered searches in the advanced detector era}

\author{Alexander Dietz}
\affiliation{Department of Physics and Astronomy, University of Mississippi, Oxford, Mississippi 38677, USA}
\author{Nickolas Fotopoulos, Leo Singer}
\affiliation{LIGO Laboratory, California Institute of Technology, Pasadena, California 91125, USA}
\author{Curt Cutler}
\affiliation{Jet Propulsion Laboratory, California Institute of Technology, Pasadena, California 91109, USA}

\date{\today}

\begin{abstract}
Short, hard, \acp{GRB} are believed to originate from the coalescence of two \acp{NS} or a \ac{NS} and a \ac{BH}. If this scenario is correct, then short~\acp{GRB} will be accompanied by the emission of strong \acp{GW}, detectable by \ac{GW} observatories such as \acs{LIGO}, Virgo, KAGRA, and \acs{LIGO}\nobreakdashes--India. As compared with blind, all\nobreakdashes-sky, all\nobreakdashes-time \ac{GW} searches, externally triggered searches for \ac{GW} counterparts to short~\acp{GRB} have the advantages of both significantly reduced detection threshold due to known time and sky location and enhanced \ac{GW} amplitude because of face\nobreakdashes-on orientation. Based on the distribution of \aclp{SNR} in candidate \acl{CBC} events in the most recent joint \acs{LIGO}\nobreakdashes--Virgo data, our analytic estimates, and our Monte Carlo simulations, we find an effective sensitive volume for \ac{GRB}\nobreakdashes-triggered searches that is $\approx$2~times greater than for an all\nobreakdashes-sky, all\nobreakdashes-time search. For \acs{NSNS} systems, a jet angle $\theta_j=20^\circ$, a gamma-ray satellite \acl{FOV} of 10\% of the sky, and priors with generally precessing spin, this doubles the number of \acs{NSNS}\nobreakdashes---short~\ac{GRB} and \acs{NSBH}\nobreakdashes---short~\ac{GRB} associations, to $\sim$3\nobreakdashes--4\% of all detections of \acsp{NSNS} and of \acsp{NSBH}. We also investigate the power of tests for statistical excesses in lists of subthreshold events, and show that these are unlikely to reveal a subthreshold population until finding \ac{GW} associations to short~\acp{GRB} is already routine. Finally, we provide useful formulas for calculating the prior distribution of \ac{GW} amplitudes from a \acl{CBC}, for a given \ac{GW} detector network and given sky location.
\end{abstract}

\maketitle

\section{Introduction}

We currently sit between the first and second generations of kilometer\nobreakdashes-scale, ground\nobreakdashes-based interferometric \ac{GW} detectors. The first direct detection of \ac{GW}s will very likely occur before the end of the decade.

The first detected signals will probably be from mergers of \ac{NSNS}, \ac{NSBH}, and \ac{BHBH} binaries, collectively referred to as \aclp{CBC}, or \acsp{CBC}. \ac{NSNS} and \ac{NSBH} coalescences are also likely progenitors for most short~\acp{GRB}~\citep{Kouveliotou:1993,Horvath:2002}. It is therefore natural to use detections of short~\acp{GRB} to trigger searches for \ac{GW} signatures of \acp{CBC} that occur at the same instant (to within a few seconds) and the same sky location (within the error bars). Such \ac{GRB}\nobreakdashes-triggered searches for \acp{GW} from \acp{CBC} are already being carried out~\citep{grb070201,DietzProcGRB,S5GRB}.

In this paper, we address several questions regarding \ac{GRB}\nobreakdashes-triggered \ac{CBC} searches.  We begin by reviewing recent results from such searches in Sec.~\ref{s:recent_results} and consider what \ac{GW} and \ac{GRB} detectors will be available in the advanced detector era in Sec.~\ref{s:roadmaps}. In Sec.~\ref{s:prospects}, we discuss the two most important factors in detectability of \ac{GW} counterparts of \acp{GRB}: namely, enhanced \ac{GW} amplitude due to preferentially low binary inclination, and the reduced \ac{GW} detection threshold resulting from knowledge of the \ac{GRB}'s time. Our primary results appear in Sec.~\ref{s:results}, in which we estimate the rate of coincident detections both analytically and via Monte Carlo simulations. A closely related data analysis activity has been the search for a statistical excess of high\nobreakdashes-\ac{SNR}, but subthreshold, \ac{CBC} candidate events coincident with short~\acp{GRB}.  In Sec.~\ref{s:stat_excess} we predict the science yield from searching for such statistical excesses, and demonstrate that the extra information will typically be negligible. Finally, in Appendix~\ref{app:responses} we derive a number of useful analytic formulas for describing a detector network's sensitivity to a \ac{CBC} at a given sky location (i.e., the location of a \ac{GRB}).

We note that rates of short~\ac{GRB}\nobreakdashes--\ac{GW} coincident detections were also estimated in recent papers by \citet{ChenHolz2012} and another by \citet{Kelleyetal2012} which appeared when this paper was almost finished. Our method is similar to both of theirs. Our conclusions are qualitatively similar to those of \citep{ChenHolz2012}, but are qualitatively different from \citep{Kelleyetal2012} due to different assumptions and approximations: \citep{Kelleyetal2012, ChenHolz2012} both assume a Gaussian distribution of outliers, while we base our calculations on the distribution of outliers observed in \ac{S6VSR23}. \citep{Kelleyetal2012} also adopt a fiducial value for the \ac{GRB} jet angle that is a factor $\sim 2$ smaller than ours. (Both values are plausible, given the uncertainties.) Because of our different assumptions about the statistics of the \ac{GW} search and the \ac{GRB} jet angle, we derive a \ac{GRB}\nobreakdashes--\ac{GW} detection rate that is $\sim 20$ times higher than theirs, which lifts it from $\sim 1/10\text{ yr}$ to $\sim 2/\text{yr}$ (i.e., from almost negligible to interesting). \citet{Nissanke:2012} perform a similar investigation but focus on the detectability of optical counterparts to \ac{GW} triggers rather than of \ac{GW} counterparts of \ac{EM} triggers. However, their simulations do encompass the targeted, \ac{EM}\nobreakdashes-triggered \ac{GW} search scenario, in their tables and figures denoted by the label \textit{Net5b}. They predict approximately the same number of \ac{GRB}\nobreakdashes--\ac{GW} detections as we do, because although they assume a steeper reduction in \ac{SNR} threshold relative to an all\nobreakdashes-sky \ac{GW} search, they also assume a smaller \ac{GRB} jet opening angle.

\subsection{Science motivation}

The detection of \acp{GW} from \acp{CBC} will have several scientific implications. The masses of the two compact objects will be determined quite accurately~\citep{Cutler:1994,FinnChernoff:1993}. With sufficiently high \ac{SNR}, the spins of these objects can also be constrained \citep{Poisson:1995ef}. These measurements, and the overall rates, will provide information on  stellar evolution \citep{0004-637X-672-1-479}. The details of the late inspiral and postmerger gravitational waveform will also inform the high-density \ac{NS} equation of state \citep{flanagan:021502,Read:2009}. Details of the merger will also permit tests of general relativity in the strong field regime \citep{Will:2005va}, tests of local Lorentz invariance \citep{Ellis2006402}, and constraints on the graviton mass \citep{PhysRevD.80.044002,Keppel:2010qu}.

A coincident short~\ac{GRB}\nobreakdashes--\ac{GW} detection would prove that at least some \acp{GRB} are indeed produced by merger events.  Furthermore, it should be possible to determine the redshift of the short~\ac{GRB}'s host galaxy, while the \acp{GW} accurately encode the distance to the binary. It has been shown that ten short~\ac{GRB} with redshift measurements could constrain  $H_0$ to within 2\% assuming a \ac{GRB} jet angle of $20^\circ$ \citep{Dalal:2006qt}.

\subsection{Recent results}
\label{s:recent_results}

To date, two types of searches for \ac{CBC} signals associated with short~\acp{GRB} have been executed: single\nobreakdashes-event targeted analyses for short~\acp{GRB} associated with very nearby galaxies \citep[\acp{GRB}~070201 and 051103;][]{grb070201,2012ApJ...755....2A}, and analyses covering all short~\acp{GRB} during \ac{LIGO} and Virgo data\nobreakdashes-taking epochs \citep{S5GRB,S6GRB}. None of these analyses found a significant \ac{GW} candidate, but the results were used to establish lower limits on the distances, assuming \ac{CBC} progenitors.

\acp{GRB}~070201 and 051103 had localizations that significantly overlapped with the galaxies  M31 \citep[Andromeda, 770\,kpc away;][]{2007GCN-6098, 2007GCN-6103} and M81 \citep[3.6\,Mpc away;][]{MNR:MNR16118}, respectively.  However searches in contemporaneous GW data were able to exclude CBCs as the source of those bursts~ \citep{grb070201,2012ApJ...755....2A}.

The 22 short~\acp{GRB} that occurred during \ac{S5VSR1} were followed up with \ac{CBC} searches in the \ac{GW} data, using analysis methods similar to those for \acp{GRB}~070201 and 051103~\citep{S5GRB}. The 26 short~\acp{GRB} that occurred during \ac{S6VSR23} were also followed up \citep{S6GRB}, using an improved, coherent analysis strategy \citep{Harry:2010fr}.  No coincidences were found, and lower distance limits on putative \ac{CBC} counterparts were established.  In the more sensitive, \ac{S6VSR23} search, the median 90\% confidence lower limits for \ac{NSNS} and \ac{NSBH} binaries were $16$ and $28$\,Mpc, respectively.  Both analyses included a test for a subthreshold population excess. For \ac{S5VSR1}, a  Mann\nobreakdashes--Whitney $U$ test~\citep{MannWhitney} was used, while a binomial test was used for \ac{S6VSR23}.  In both cases, the subthreshold populations were found to be consistent with the background.

\section{Detector network roadmap}
\label{s:roadmaps}

To make sensible predictions of the outcome of future \ac{GRB}\nobreakdashes-triggered \ac{GW} searches, one needs to know what \ac{GRB} and \ac{GW} detectors might be operating in the next decade.

\subsection{\ac{GW} detector network roadmap}

The U.S. \ac{LIGO} \citep{Abbott:2007kv} has recently completed a one\nobreakdashes-year data\nobreakdashes-taking period between July 2009 and October 2010, in coincidence with the French\nobreakdashes--Italian Virgo detector \citep{Acernese:2008zzf}.

\ac{LIGO} is currently upgrading to its advanced detector configurations \citep{Harry:2010zz}, with the goal of increasing the sensitivity gradually to a factor of $\sim$10 compared to the initial configuration and extending seismic\nobreakdashes-limited sensitivity to lower frequencies. More recently, the U.S. National Science Board has authorized one \ac{LIGO} detector to be moved to India in order to vastly expand the worldwide detector network's sky localization capabilities. The earlier attempts to do the same in Australia have been formally abandoned.

Virgo is upgrading to the Advanced Virgo configuration \citep{Acernese:2008zzf}, similar to Advanced \ac{LIGO} in optical layout and sensitivity. The start of the data\nobreakdashes-taking period with the advanced detectors is foreseen in $\sim$2015.

GEO600 is a British-German detector with 600\,m arms and advanced optical configurations \citep{Grote:2010zz}. GEO600's next\nobreakdashes-generation configuration will be GEO\nobreakdashes-HF with a focus on high\nobreakdashes-frequency sensitivity \citep{2006CQGra..23S.207W}. GEO\nobreakdashes-HF's high\nobreakdashes-frequency sensitivity will be the best in the world, which will be useful in parameter estimation \citep{Read:2009}. However, low\nobreakdashes-frequency sensitivity is more important for \ac{CBC} detection and GEO600's relatively short arms put it at a significant disadvantage. It will likely not be used for detection searches.

Construction has begun on KAGRA (KAmioka GRAvitational\nobreakdashes-wave observatory, formerly LCGT) in Japan~\citep{0264-9381-27-8-084004}, which should reach its design sensitivity in 2018. It has 3\,km\nobreakdashes-long arms constructed underground and uses cryogenically cooled sapphire mirrors for test masses.  The final detector is expected to detect a \ac{NSNS} system at a distance of 240\,Mpc with \ac{SNR}=10~\citep{Uchiyama2004}.

\subsection{\ac{GRB} detector network roadmap}

During the last \ac{S6VSR23} science run, the triggers came mostly from the \textit{Swift} and \textit{Fermi} missions, with a few from the \ac{IPN}. \ac{IPN} detected most of the \textit{Swift}/\textit{Fermi} triggers too, but with a much poorer sky localization because only triangulation methods can be used by \ac{IPN}.

\ac{IPN} has unfortunately lost its primary funding, but is nonetheless expected to operate during 2015\nobreakdashes--2020, if perhaps with a smaller number of satellites, still detecting \acp{GRB} at these times but with a lower rate~\citep{PrivateCommunicationHurley2011}. \textit{Swift} might operate for another 5 years or even longer, since it has no expendables and the spacecraft is in good shape~\citep{PrivateCommunicationBurrows2011}, but the operation depends on NASA funding. The \textit{Fermi} instruments might also operate until~$\sim$$2018$. The \ac{GBM} instrument on \textit{Fermi} achieves instantaneous sky coverage of about 70\% or 8.8~sr~\citep[30\% of the sky being occulted by the Earth at the altitude of Fermi's orbit;][]{wilson-hodge-2012}, but \acp{GRB} detected by the \ac{GBM} alone are very poorly localized.

\textit{Lobster} is a proposed NASA mission similar to \textit{Swift} in strategy, with a wide-field X-ray imager (WFI), narrow-field followup IR telescope (IRT), and slewing apparatus to point the latter \citep{lobster}. WFI is more sensitive than the \textit{Swift} \ac{BAT}, but has a smaller \ac{FOV} at 0.5\,sr.

The French\nobreakdashes--Chinese \ac{SVOM} mission is targeted at a broader scientific target, including answering questions related to \acp{GRB}, cosmology, and fundamental physics \citep{Paul:2011ii}. Its main instrument, ECLAIR, is a coded-aperture telescope aimed at a broad energy range of 4\nobreakdashes--250\,keV, with a \ac{FOV} comparable to that of \textit{Swift}. The effective detection area is also close to that of \textit{Swift}, resulting in an expected $70$\nobreakdashes--$90$ \acp{GRB} per year, of which $20$\nobreakdashes--$25$\% could be short \acp{GRB}. As the ECLAIR telescopes are more sensitive to lower energies compared to \ac{BAT}, the extended emission and afterglows of \acp{GRB} can be observed deeper, resulting in improved \ac{GRB} locations, and hence in a larger number ($\sim$50\%) of \acp{GRB} with redshift measurements~\citep{PrivateCommunicationBasa2011}. The anticipated launch date is $\sim$2015\nobreakdashes--2020.

The South Korean-led \ac{UFFO-P} mission intends to catch the rise of \acp{GRB} \citep{2012arXiv1207.5759G}. It has been constructed and is anticipated to launch in June 2013. It carries a coded-aperture burst alert telescope similar to \textit{Swift}'s \ac{BAT}, sensitive from 15\nobreakdashes--200~keV and with a \ac{FOV} of $\sim$2~sr. \ac{UFFO-P}'s headlining feature is that it can repoint in response to a trigger in $\sim 1$~s using its \ac{SMT}, which is a substantial improvement in response time over \textit{Swift}'s $\sim$1~minute to slew the whole spacecraft. Though \ac{UFFO-P} has a small collecting area and only a small optical telescope for followup, this pathfinder mission already has some discovery potential. The conceived \textit{UFFO-100} mission will increase collecting area, replace the \ac{SMT} with a still faster MEMS micromirror array to redirect its optical path, and add an NIR camera, with the goal of gathering a statistically significant population of rising \acp{GRB}.

In conclusion, several missions are expected to operate during the advanced detector area, which are either already operating (like \textit{Swift}, \textit{Fermi} and \ac{IPN}3), are in development (like \ac{SVOM} and \ac{UFFO-P}), or planned (like \textit{Lobster} and \textit{UFFO-100}). However, given the uncertainty in how many of these missions will ultimately fly, throughout this paper we will assume that during the advanced \ac{GW} detector era the effective coverage of the combined \ac{GRB} detector network will be approximately that of \textit{Swift}'s \ac{BAT} alone, 1.4~sr~\citep{Barthelmy:2005hs} or about one\nobreakdashes-tenth of the sky.

\section{Detection prospects}
\label{s:prospects}

\subsection{Collimation}
\label{sec:collimation}

\acp{GRB} show strong evidence for collimated, relativistic outflow along a jet. Assuming that the jet is roughly conical, its size is described by the jet angle $\theta_j$, from the center to its outer edge.  We define $f_\mathrm{b}$ to be the fraction of the sky into which gamma rays are launched.  For a CBC that emits a single jet, this is
$f_\mathrm{b}=(1-\cos{\theta_j}) / 2$.   If the CBC emits two jets, presumably in opposite directions, then this fraction doubles to $f_\mathrm{b}=(1-\cos{\theta_j})$.
Collimation reduces the number of CBCs that are observable as short~\ac{GRB} events,  since the observer only sees the \acp{GRB}  for which the Earth lies within the jet.

The Lorentz factor $\Gamma$ of the beam decreases as it sweeps up external material, and at the point where $\Gamma$ reaches $\sim 1/\theta_j$, the flux decay abruptly steepens due to special relativistic effects.   The beaming half-angle can be determined for a given \ac{GRB} by the time of this ``jet break.''  However the rapid decay of the late-time lightcurves of short~\acp{GRB} makes the estimation of $\theta_j$ difficult.  \citet{Grupe:2006uc} places a lower limit of $\theta_j \gtrsim 25^\circ$ while \citet{Burrows:2006ar} infers the value of $\theta_j$ to be in the range $4$--$8^\circ$. \citet{Goldstein:2011ju} suggest a value in the range between $40^\circ$ and $90^\circ$. \citet{2012arXiv1204.5475F} find $3$--$8^\circ$ in a recent short~\ac{GRB}. Simulations of \ac{NSBH} mergers indicate a range of $\theta_j\simeq30$--$50^\circ$ for binaries of moderate spin while finding $5$--$10^\circ$ for near-extremal spinning systems \citep{Foucart:2010eq}. In this paper we will adopt $\theta_j = 20^\circ$ as a fiducial value when quoting results; however our analytic estimates and Tables allow the reader to trivially convert 
the results to other values of $\theta_j$.

The \ac{GRB} beaming angle is presumably not a universal constant, but has a distribution.  While our simulations assumed that all \acp{GRB} had the same value for $f_\mathrm{b}$, our results on rates in this paper are approximately generalizable by simply replacing $f_\mathrm{b}$ by its average value $\langle f_\mathrm{b} \rangle$.  We emphasize that $\langle f_\mathrm{b} \rangle$ refers to the average over all  short~\acp{GRB}, not just the detected ones, since the detected population depends on selection effects. (E.g., the detected population is biased towards \acp{GRB} with larger values of $f_\mathrm{b}$, at fixed flux.)

\subsection{Reduced search space}
\label{sec:reduced}

The search for a \ac{GW} \ac{CBC} signal triggered on an \ac{EM} counterpart has sensitivity advantages over an all\nobreakdashes-sky search. Semi\nobreakdashes-analytic calculations and numerical simulations predict that the majority of the \ac{NS} matter is accreted within milliseconds to seconds. This has guided \ac{GRB}\nobreakdashes-triggered \ac{CBC} detection efforts to search only a $[-5, 1]$~s `on\nobreakdashes-source window' surrounding a \ac{GRB} trigger to account for up to a 5\nobreakdashes-second \ac{GRB}\nobreakdashes--\ac{GW} delay and up to 1~s of uncertainty in the \ac{GRB} \ac{TOA} \citep[see sections 2.2 and 5.1 of][for references]{S6GRB}. There are further possible reductions in the searched parameter space due to the known sky location and even by restricting to the space of \ac{CBC} parameters that allow for tidal disruption outside the innermost stable circular orbit, but in this paper we will neglect them, as they have much less impact than the reduction in observation time. 

In this section, we are interested in estimating the reduction in \ac{SNR} threshold in going from an all-sky search to a \ac{GRB}\nobreakdashes-triggered search while keeping constant the \ac{FAP} at the detection threshold. The first detection is likely to be held to a high standard of $\ac{FAP} \lesssim 10^{-6}$, but once detections are routine, the threshold should be lessened considerably. Throughout this paper, we assume that $\ac{FAP} \lesssim 10^{-4}$ is required.

For 20 short~\acp{GRB} per year of livetime ($T_\mathrm{all}=1\,\text{yr}$), the observation time for \acp{GW}  is $T_\mathrm{grb}=120\,\text{s}$. Assuming that the searches are comparably effective at background rejection, the \ac{FAR} at a given \ac{SNR} should be the same, but the \ac{FAR} at a given \ac{SNR}, $\acs{FAP}(\rho) = T\, \acs{FAR}(\rho)$, is reduced by a factor
\begin{equation*}
\frac{\acs{FAP}_\mathrm{grb}(\rho)}{\acs{FAP}_\mathrm{all}(\rho)} = \frac{T_\mathrm{grb}}{T_\mathrm{all}} \approx 4\times10^{-6} \,.
\end{equation*}

We estimated the \ac{SNR} threshold for a \ac{GRB}\nobreakdashes-triggered search using the background \ac{SNR} distribution from the \ac{S6VSR23} all\nobreakdashes-sky search~\citep{LIGO:2011nz}. The all\nobreakdashes-sky search was a \textit{coincident} search, in which matched filtering and thresholding were performed on individual detectors and spurious triggers were vetoed by demanding consistent \acp{TOA} in multiple detectors. For a targeted, \ac{GRB}\nobreakdashes-triggered search, a \textit{coherent} search is possible in which thresholding is done on the suitably time\nobreakdashes-delayed and summed \ac{SNR} for the whole network~\citep{Harry:2010fr}. False\nobreakdashes-positive rejection is aided by tests on null streams, which are special linear combinations of the detectors that are insensitive to \acp{GW}. Coherent searches are less feasible for all\nobreakdashes-sky searches because each sky location requires unique time delays. Since in Gaussian noise a coincident search with a two\nobreakdashes-detector network has the same statistics as a coherent search with any network of (more than one) detectors~\citep{Harry:2010fr}, we can extrapolate the threshold for a targeted, coherent search from the statistics of an all\nobreakdashes-sky, coincident search.

For the \ac{S6VSR23} all\nobreakdashes-sky search, Fig.~3 of \citep[][data available online at \url{https://dcc.ligo.org/LIGO-P1100034-v19/public}]{LIGO:2011nz} gives the \ac{FAR} as a function of $\rho_c$, a $\chi^2$\nobreakdashes-weighted quadrature sum of the \ac{SNR} in all of the detectors that has been found to be a useful detection statistic. We assume that this $\rho_\text{c}$ is equivalent to the network coherent \ac{SNR} $\rho$, as argued above. From this Figure, we find that during \ac{S6VSR23}, $\ac{FAR}=10^{-4}~\text{yr}^{-1}$ when $\rho_c = 11.3$. When $\ac{FAR}=10^{-4} / 120~\text{s}$, $\rho_c = 9$. From this, we take $\rho_{\mathrm{th},\mathrm{all}}=11.3$ as the \ac{SNR} threshold for an all\nobreakdashes-sky search and $\rho_{\mathrm{th},\mathrm{grb}}=9$ as the \ac{SNR} threshold for a \ac{GRB}\nobreakdashes-triggered search. A triggered search can see $11.3/9$ times farther than the all\nobreakdashes-sky search. On the other hand, as we will see, the increased range is somewhat offset by jet collimation and the limited \acp{FOV} of high\nobreakdashes-energy satellites.

\section{Results}
\label{s:results}

\subsection{Analytic estimates}
\label{sec:analytic}

Here we provide some simple estimates of the \ac{CBC} detection rate we expect for triggered searches, compared to the \ac{CBC} detection rate for untriggered, all\nobreakdashes-sky searches. The method is the same as in \citet{KochanekPiran1993}, but our inputs and conclusions are different. We will assume that short~\acp{GRB} come \textit{only} from \acp{CBC}, and denote by $f_\text{GC}$ the fraction of \acp{CBC} that produce short~\acp{GRB}.  For the \acp{CBC} that produce bursts, let $f_\text{b} $ be the average solid angle into which gamma rays are launched. Again, the average is over the whole \ac{GRB} population, not just population of detected \acp{GRB}  (which is biased towards \acp{GRB} with wider beams at fixed flux). For any \ac{CBC} that produces two beams (presumably in opposite directions), we consider $f_\text{b}$ for \textit{that} burst to be the sum of the solid angles for the two beams. Finally, we define $\bar S$ to be the average fraction of the sky that is ``covered'' by the then-existing \ac{GRB} detector network. For the advanced \ac{GW} detectors, most \acp{CBC} will be detected at distances $\lesssim 200\,$Mpc, while most detected short~\acp{GRB} are much farther away ($D \sim 1\,$Gpc), so the \ac{GRB} accompanying an observed  \ac{CBC} should be detectable as long as the Earth lies within the beam, and the source is within the telescope's \ac{FOV}. Then the fraction $F$ of \acp{CBC} for which we detect a \ac{GRB} is
\begin{equation}
F = f_\text{GC}\, \bar S \,   f_\text{b}
\end{equation}
We shall adopt $F = 6\times 10^{-3}$ as a reasonable fiducial number. This would correspond, e.g.,  to $\bar S = 0.1$, and all CBCs emitting two jets of $\theta_j = 20^{\circ}$ each.

As explained above, by looking for \ac{NS} binaries that are roughly coincident in time with observed short~\acp{GRB}, we increase the sensitivity of the search at fixed \ac{FAR}. Let $\rho_{\mathrm{th},\mathrm{all}}/\rho_{\mathrm{th},\mathrm{grb}}$ be the ratio of the \ac{GW} detection thresholds for blind and \ac{GRB}\nobreakdashes-triggered searches, respectively. As explained in Sec.~\ref{sec:reduced}, we adopt as a fiducial value $\rho_{\mathrm{th},\mathrm{all}}/\rho_{\mathrm{th},\mathrm{grb}} \approx 11.3/9.0 \approx 1.25$ .

The final factor that  we need arises because the mergers that we see in gamma rays are presumably the ones for which we are viewing the binary nearly face on (because the gamma rays are presumably beamed perpendicular to the orbital plane.) The quadrupolar pattern of the emitted \acp{GW} is also strongest along the direction perpendicular to the orbital plane; the amplitude of $h$ on the axis is $\approx 1.51$ times stronger than the isotropic detection-averaged value. (It is well know that that rms enhancement is $\sqrt{5/2}=1.58$, but what is relevant for determining rates is the cube root of mean-cubed enhancement, which is $1.51$.) Although $1.51$ is the enhancement factor of an optimally oriented binary, we show in Appendix~\ref{app:responses} that as long as the beam half-angle is $< 25^{\circ}$, then using $1.51$ is at most a $\sim 5\%$ overestimate, which is acceptable for our purposes.

For Advanced \ac{LIGO}, we will be detecting \acp{CBC} in the range $\sim 50\,$Mpc to $\sim 500\,$Mpc. At these distances, we can safely approximate spatial geometry as Euclidean and the density of mergers as uniform. Let $\Ndot_\mathrm{all}$ be the \ac{CBC} detection rate of blind, all\nobreakdashes-sky advanced \ac{GW} detector searches, and let  $\Ndot_\mathrm{grb}$ be the rate for triggered searches.  What is the ratio $\Ndot_\text{grb}/\Ndot_\text{all}$?  The volume in which \ac{GRB}\nobreakdashes-triggered \acp{CBC} are detectable is larger than for the blind case by the factor $(1.51 \rho_{\mathrm{th},\mathrm{all}}/\rho_{\mathrm{th},\mathrm{grb}})^3$, but recall that only a fraction $F$ of the \acp{CBC} emit detectable gamma rays. Thus,
\begin{equation}
\frac{\Ndot_\mathrm{grb}}{\Ndot_\mathrm{all}} = \left(1.51 \frac{\rho_{\mathrm{th},\mathrm{all}}}{\rho_{\mathrm{th},\mathrm{grb}}} \right)^3 F \sim 0.041.
\end{equation}
In other words, of the first $\sim 25$ detections, we would expect only one to come from the \ac{GRB}\nobreakdashes-triggered pipeline; it is correspondingly unlikely that the very first \ac{GW} detection of a \ac{CBC} will result from a \ac{GRB} trigger.  This is the same conclusion reached in \citet{ChenHolz2012} and \citet{Kelleyetal2012}.

Of the \ac{CBC} detections associated with short~\acp{GRB}, a fraction $(\rho_{\mathrm{th},\mathrm{all}}/\rho_{\mathrm{th},\mathrm{grb}})^3 \sim 0.5$ will be strong enough to be detectable even without the \ac{GRB} trigger. So the \textit{increase} in the rate of \ac{GW} detections, thanks to \ac{GRB} triggers, will be $\sim 2\%$. This is a small increase in \ac{CBC} detections, but, given the extra information to be gleaned from the \ac{EM} counterpart, a non-negligible one. Note that this $2\%$ increase is about $2$ times higher than the fiducial increase reported in \citet{Kelleyetal2012}. The difference comes mostly from two factors: i)~a factor $\sim 4$\nobreakdashes--$5$ from our use of non-Gaussian statistics, and ii)~a factor $\sim 6$\nobreakdashes--$7$ from our wider fiducial beam angle of $20^{\circ}$ vs.\ their $11^{\circ}$, combined with our assumption of two opposite\nobreakdashes-pointing jets per \ac{CBC}.

\subsection{Effective sensitive volumes}

For a given detector network and mass pair, we would like to compute the
relative detection capability of a targeted short~\ac{GRB} search compared to an
all-sky search. 
Following \citet{FinnChernoff:1993}, but generalizing from a single detector to a network,
we define an effective sensitive volume
$V_\mathrm{sens}$, which can be multiplied by a constant rate density $\mathcal{R}$
to obtain the total detection rate $\Ndot = \mathcal{R} V_\mathrm{sens}$:
%
%
\begin{align}
V_\mathrm{sens} &= \left\langle \int \Theta(\rho - \rho_\mathrm{th}) dV\nonumber \right\rangle_{\psi,\iota}\\
&= \int d\Omega \left\langle  \int_0^{D_\mathrm{th}} r^2 dr \right\rangle_{\psi,\iota}\nonumber\\
&= \frac{4\pi}{3} \left\langle D^3_\mathrm{th} \right\rangle_{\psi,\iota,\delta,\alpha} \,,
\label{eq:sens_distance}
\end{align}
where $\Theta$ is the Heaviside step function, $(\alpha, \delta)$ are the source's right ascension and declination, and $D_\mathrm{th}$ is the distance
at which the network registers \ac{SNR} above its threshold $\rho_\mathrm{th}$ for a given
source type. $D_\mathrm{th}$ depends on the full specification of a waveform plus the detector
noise spectrum and the various projection angles. Eq.~\ref{eq:sens_distance}
is quite convenient for Monte Carlo integration.
See Appendix~\ref{app:responses} for network \ac{SNR} expressions and analytical
evaluation of Eq.~\ref{eq:sens_distance} in the nonspinning case.

To take into account the imperfect duty cycle of the detectors, along with
our requirement that at least two detectors be ``on'' to claim a detection, it is useful to define
a \textit{double detection volume} (2DV), in close analogy to the triple detection rate (3DR) of
\citet{0264-9381-28-12-125023}. (Our formulation generalizes the treatment in \citet{0264-9381-28-12-125023} since it applies to spinning binaries and removes the 
 assumption of equal detector responses.) 
%
For a set of detectors $\mathcal{N}$,
\begin{equation}
\text{2DV} = \sum_{\mathcal{X}\in \mathcal{P}_{\geq 2}(\mathcal{N})} f^{|\mathcal{X}|} (1 - f)^{|\mathcal{N}| - |\mathcal{X}|} V^\mathcal{X}_\text{sens} \,,
\label{eq:2DV}
\end{equation}
where $\mathcal{P}_{\geq 2}(\mathcal{N})$ is the set of subsets of $\mathcal{N}$ with size 2 or greater and $|\mathcal{X}|$ denotes the size of set $\mathcal{X}$. $f$ is the duty cycle, which we take to be the same for all detectors.

\subsection{Results from Simulations}

We evaluate the 2DV (equation~\ref{eq:2DV}) via Monte Carlo integration, 
randomly drawing
the sky location and the polarization angles, for both \ac{NSNS}
(1.4--1.4\,$M_\odot$) and \ac{NSBH} (1.4--10\,$M_\odot$) pairs.
We use the Taylor T4 waveform \citep{Boyle:2008ge}, including general spin
precession, to determine expected signal amplitudes and thus $D_\mathrm{th}$.
We match the distributions of spin amplitude and spin orientation used
in the recent GRB~051103 analysis \citep{2012ApJ...755....2A} with NS
dimensionless spin drawn uniformly from $[0, 0.4]$, BH spin drawn
uniformly from $[0, 0.98]$, and uniform spin orientations subject to a cut on the
tilt angle (the angle between the \ac{BH} spin axis and \ac{NS} orbital axis) to be
$<$60$^\circ$. We also adopt their rigid rotation procedure so that it is the
total angular momentum that is uniformly distributed within the cone of
$\theta_j$ rather than the orbital angular
momentum. Using different values for $\theta_j$, we compute the
relative improvement in effective sensitive volume 2DV of a triggered search
(restricted inclination angle, coherent threshold $\rho_{\mathrm{th},\mathrm{grb}}=9.0$) compared to an all\nobreakdashes-sky search
(unrestricted inclination, coincident threshold $\rho_{\mathrm{th},\mathrm{all}}=11.3$). 

The 2DVs have been defined such that the detection rate for the all\nobreakdashes-sky search
is $\Ndot_\mathrm{all}=\mathcal{R}\,V_{90^\circ}^{11.3}$, where
$\mathcal{R}$ is the merger rate density in the local Universe and
$V_{90^\circ}^{11.3}$ the 2DV assuming a coherent \ac{SNR} threshold of 11.3 and a
source population whose inclination is unrestricted
($\theta_j=90^\circ$). Similarly, the number of \acp{GRB} associated
with a \ac{GW} detection is given by $\Ndot_\mathrm{grb} =
\bar S\, f_\mathrm{b} \,\mathcal{R}\,V_{\theta_j}^{9}$.
The rate for \acp{GRB} independently detected by both searches (i.e., with GW network SNR $> 11.3$)
is $\Ndot_\mathrm{both} =
\bar S\, f_\mathrm{b} \,\mathcal{R}\,V_{\theta_j}^{11.3}$.
Our computed sensitive volumes are shown in Table~\ref{tab:MCresults}.
Each value of $V_{\theta_j}^{9}$ represents $10^6$ simulations, while 
the $V_{\theta_j}^{11.3}$ values are simply rescaled from
$V_{\theta_j}^{9}$.

\begin{figure}[h!]
\centering
\includegraphics[width=3in,angle=0]{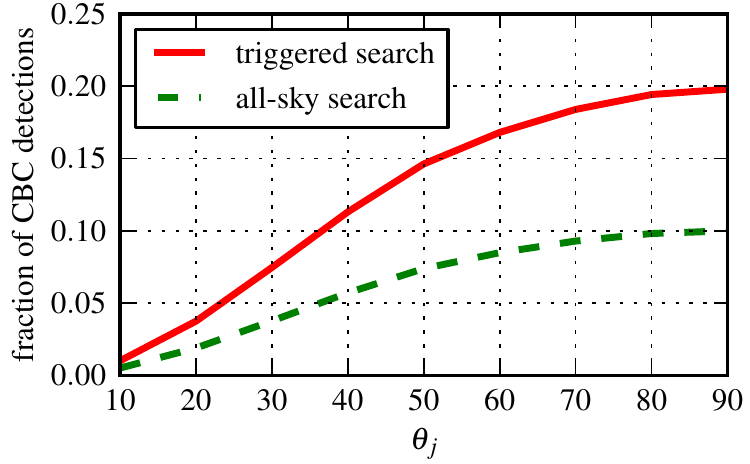}
\caption{Fractional number of additional \ac{CBC} detections contributed by an
externally triggered search (solid) and fraction of all\nobreakdashes-sky \ac{CBC} detections
predicted to be in coincidence with short~\ac{GRB} observations without a
special triggered search (dashed). \ac{NSNS} and \ac{NSBH} ratios are equal within
statistical uncertainty, so only \ac{NSNS} results are plotted.}
\label{fig:response_opening}
\end{figure}

Under the assumption that all \acp{CBC} produce
short~\acp{GRB} and vice versa, we translate sensitive volumes to detection rates in
Table~\ref{tab:ABSresults} and Fig.~\ref{fig:response_opening}.
We take values of $0.0198\,L_{10}/\text{Mpc}^3$ and ``realistic'' coalescence rates
of $6\times10^{-5}\,L^{-1}_{10}\,\text{yr}^{-1}$ for \ac{NSNS} sources
and $2\times10^{-6}\,L^{-1}_{10}\,\text{yr}^{-1}$ for \ac{NSBH}, as
used in \citet{Abadie:2010cf}. The coalescence rates are uncertain by two
orders of magnitude, but our relative volumes have much smaller errors, so we
show two or more significant figures not to reflect
uncertainty but rather to allow for relative comparisons. We have assumed an effective
\ac{FOV} of $\bar S=0.1$, independent duty cycle factors of $f=0.8$, and a network
of \ac{LIGO}\nobreakdashes--Hanford, \ac{LIGO}\nobreakdashes--Livingston, Virgo, KAGRA~\citep[final orientation via ][]{PrivateCommunicationTakashi2012}, and \ac{LIGO}\nobreakdashes--India~\citep[HLVKI; orientation guess via ][]{0264-9381-28-12-125023}. For all \ac{GW} detectors, we used publicly available projections of noise spectra at full design sensitivities \citep{aLIGOPSD,aVirgoPSD,KAGRAPSD}.
However, we started \ac{SNR} integration
from a lower frequency limit of 40\,Hz in order to fit waveforms within the memory limits of the bulk of
computers available to us, enabling much greater parallelism, at some cost in
realism. 
 These assumptions lead a total annual detection rate of 61 \ac{NSNS} signals and 18 \ac{NSBH} signals for the all-sky search. 

Assuming an effective \ac{FOV} of $\bar S=0.1$ for the GRB detectors, and that every \ac{CBC} produces two opposite jets, each with a fiducial jet angle of $\theta_j=20^\circ$, we find that 1.8\% of \acp{CBC} detected by the all\nobreakdashes-sky search would be coincident with observed short~\acp{GRB}. Adding the triggered search increases the number of coincidences to $3.8$\%. Clearly, our Monte Carlo results are in very close agreement with the analytic estimates in Sec.~\ref{sec:analytic}.

\begin{table}[h]
\caption{\label{tab:MCresults}Double detection volume (2DV) in Gpc$^3$.}
\begin{ruledtabular}
\begin{tabular}{ddddd}
& \multicolumn{2}{c}{\ac{NSNS}} & \multicolumn{2}{c}{\ac{NSBH}} \\
\cmidrule[0.5pt](lr{0.2em}){2-3} \cmidrule[0.5pt](lr{0.2em}){4-5} \\[-8pt]
\theta_j & \multicolumn{1}{c}{$V_{\theta_j}^{9}$} & \multicolumn{1}{c}{$V_{\theta_j}^{11.3}$} &\multicolumn{1}{c}{$V_{\theta_j}^{9}$} & \multicolumn{1}{c}{$V_{\theta_j}^{11.3}$}  \\
\colrule
10 & 0.31 & 0.16 & 2.5 & 1.3 \\
20 & 0.29 & 0.15 & 2.4 & 1.2 \\
30 & 0.26 & 0.13 & 2.1 & 1.1 \\
40 & 0.23 & 0.11 & 1.9 & 0.95 \\
50 & 0.19 & 0.097 & 1.6 & 0.81 \\
60 & 0.16 & 0.080 & 1.3 & 0.68 \\
70 & 0.13 & 0.066 & 1.1 & 0.57 \\
80 & 0.11 & 0.056 & 0.96 & 0.48 \\
90 & 0.093 & 0.047 & 0.83 & 0.42 \\
\end{tabular}
\end{ruledtabular}
\end{table}

\begin{table}[h]
\caption{\label{tab:ABSresults}Annual detection rate of short~\ac{GRB}\nobreakdashes--\ac{CBC} coincidences by the externally triggered search ($\Ndot_\text{grb}$) and by both the triggered search and the all\nobreakdashes-sky search ($\Ndot_\text{both}$). For comparison, we compute the annual detection rate for \ac{CBC} signals, with or without \ac{GRB} counterparts, by the all\nobreakdashes-sky search ($\Ndot_\text{all}$) to be $61$ and $18$ for \ac{NSNS} and \ac{NSBH} sources, respectively.}
\begin{ruledtabular}
\begin{tabular}{ddddd}
& \multicolumn{2}{c}{\ac{NSNS}} & \multicolumn{2}{c}{\ac{NSBH}} \\
\cmidrule[0.5pt](lr{0.2em}){2-3} \cmidrule[0.5pt](lr{0.2em}){4-5} \\[-8pt]
\theta_j & \multicolumn{1}{c}{$\Ndot_\text{grb}$}  &\multicolumn{1}{c}{$\Ndot_\text{both}$} & \multicolumn{1}{c}{$\Ndot_\text{grb}$} & \multicolumn{1}{c}{$\Ndot_\text{both}$} \\
\colrule
10 & 0.62 & 0.31 & 0.16 & 0.083 \\
20 & 2.3 & 1.2 & 0.61 & 0.31 \\
30 & 4.6 & 2.3 & 1.2 & 0.62 \\
40 & 6.9 & 3.5 & 1.9 & 0.95 \\
50 & 8.9 & 4.5 & 2.5 & 1.2 \\
60 & 10 & 5.2 & 2.9 & 1.5 \\
70 & 11 & 5.7 & 3.2 & 1.6 \\
80 & 12 & 6.0 & 3.4 & 1.7 \\
90 & 12 & 6.1 & 3.6 & 1.8 \\
\end{tabular}
\end{ruledtabular}
\end{table}

\section{Sub-threshold statistical excess tests}
\label{s:stat_excess}

In a calendar year, there will be some number of searches for \acp{GW} associated with short~\acp{GRB}. Some of these may result in individual detections, meaning that there will be events that survive all vetoes and have SNR above some theshold $\rho_1$ for confident single-source detection. However one can also consider some lower threshold $\rho_\text{e}$, and look for a statistical excess of events with \ac{SNR} between $\rho_\text{e}$ and $\rho_1$. What science can be gleaned from those excess events?

We begin by providing a Bayesian perspective on this question. First, setting any threshold and discarding events below that threshold amounts to ``throwing away'' data, and in the ideal case of arbitrary computing power and a perfectly known distribution of the detection statistic, discarding data weakens the analysis. So in the ideal case, $\rho_\text{e}$ would be set very low. However, in practice, finite resources, an imperfectly known distribution of outliers, and diminishing infomation returns (for lower $\rho_\text{e}$) all will push data analysts to a value of $\rho_\text{e}$ not too far below $\rho_1$. Probably the most important information encoded in the excess, sub-$\rho_1$ events is an improved estimate of the event rate ${\cal R}$ (e.g., in units of ${\rm Mpc}^{-3}{\rm yr}^{-1}$). As emphasized in \citet{2012arXiv1206.3461M}, a proper Bayesian analysis takes into account the value of $\rho_\text{e}$, and so always gives an unbiased estimate of ${\cal R}$. Including more events by lowering $\rho_\text{e}$ just ``shrinks the error bars.''

Next we provide a crude, back-of-the-envelope argument why, in the regime of rare detection, there will generally not be much information in the sub-threshold excess.  In the regime of rare detection, the unique loudest event will have \ac{SNR} $\sim 10$. There are 8 times as many events twice as far away, so we would expect an order of $8$ true events with $5<\text{SNR}<10$. However the number of background events with SNR between $5$ and $10$ will be of order $10^{-4}$ (the \ac{FAP} at $\text{SNR} \sim 10$) times $e^{-25/2}/e^{-100/2}$, or $\sim 10^{12}$ (assuming a Gaussian distribution of noise events). The standard deviation in this expected number of events is $10^6$, which swamps the true excess. Of course, this calculation was for one (nonoptimized) value of $\rho_\text{e}$, but we believe that the general conclusion is robust: as $\rho_\text{e}$ is decreased below $\rho_1$, the background rate increases far faster than the rate of true events, so we expect relatively little extra information from the sub\nobreakdashes-threshold events.

In the next subsection we consider two non-Bayesian statistical excess tests currently used in \ac{LIGO} searches, and show through simulations that, indeed, the sub-$\rho_1$ excess typically contains rather little useful information.  We believe that a proper Bayesian analysis the sub-$\rho_1$ events would lead to the same qualitative conclusion, but we leave that calculation for future work.

\subsection{Results from simulations}

In past \ac{LIGO} searches, two statistical excess tests
were employed: the Mann\nobreakdashes--Whitney\nobreakdashes--Wilcoxon $U$ test and later the binomial test
\citep{S5GRB,S6GRB}.  However we are not aware of any
systematic comparison of their detection power, so we compare them here.

\begin{figure}[t]
\centering
\includegraphics{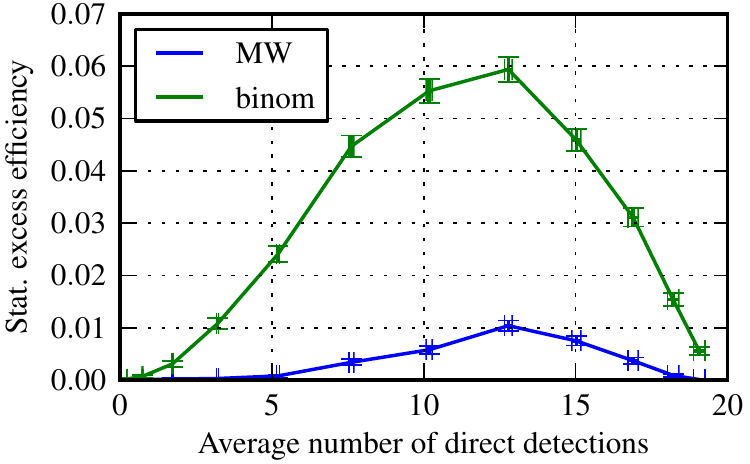}
\caption{Fraction of simulations in which the $U$ and binomial tests
detected a statistical excess in sub-threshold candidates vs the rate of
direct detections.}
\label{fig:stat_excess}
\end{figure}
We ran a Monte Carlo simulation in which each trial drew 20 \ac{GRB} on-source
candidates and, for each one, $10^6$ off-source trial loudest candidates. Off-source
candidates were drawn from the
\ac{S6VSR23} distribution (see Sec.~\ref{sec:reduced}). Each on-source
candidate \ac{SNR} was the maximum of an off-source trial and a draw from a
foreground $p(\rho)\propto\rho^{-4}$ distribution whose rate scaling could be
varied. For each trial, we performed individual direct detection searches on
all 20 \acp{GRB} before performing a statistical excess search on the set of
remaining nondetections. Both direct detection of individual \acp{GRB} and
statistical excess detection had thresholds set by $\ac{FAP} = 10^{-4}$.
Fig.~\ref{fig:stat_excess} shows the fraction of trials in which the $U$ and
binomial tests were able to detect an excess versus the average number of
direct detections among the $20$ simulated \acp{GRB}. While the $U$ test used
all on-source and all off-source candidates, the binomial considered only the
loudest four on-source candidates.

We see our expectations confirmed: for both tests, the probability of detecting
an excess in the sub-threshold populaton is never larger than several percent.
Of the two tests, the binomial one is the more powerful. We see both tests
become more powerful as the number of direct detections rises, until it reaches
$\sim 13$. The decline in efficiency for $> 13$ detections is an artifact of
the ``rules'' of our simulation in that there are always $20$ true events, so as
more of them exceed $\rho_1$, fewer true events are left among the
sub-threshold candidates.

\section{Summary and discussion}

Based on \ac{LIGO}\nobreakdashes--Virgo \ac{S6VSR23} data and search parameters for \ac{CBC}\nobreakdashes--short~\ac{GRB} searches, we have quantified how much deeper into the noise one can dig with knowledge of the external trigger time; we find a coherent \ac{SNR} threshold of $9.0$ versus an all\nobreakdashes-sky coincident threshold of $11.3$ for a detection \ac{FAP} of $10^{-4}$. \citep{ChenHolz2012,Kelleyetal2012} have also estimated the reduction in the \ac{SNR} threshold assuming Gaussian noise and making different choices in how to fold in \ac{EM} information. The bleakness of \citep{Kelleyetal2012} compared to our study is easily understood: if the measured distribution of SNRs falls off less steeply than a Gaussian, then reducing the search volume has a relatively larger effect on the detection threshold (at fixed \ac{FAP}), and hence on the detection rate. While the advanced detectors will not have the same distribution of high-SNR candidates \acp{SNR} as \ac{LIGO} and Virgo did in \ac{S6VSR23}, the true distribution seems unlikely to be Gaussian, and the difference between our results and \citep{ChenHolz2012,Kelleyetal2012} provides some measure of the importance of non\nobreakdashes-Gaussian backgrounds in assessing the value of triggered searches.

Folding the threshold reduction into a large Monte Carlo simulation, including the effects of short~\ac{GRB} collimation, general spin precession, and advanced \ac{GW} and \ac{GRB} detector networks, we have estimated the rate of \ac{CBC}\nobreakdashes--short~\ac{GRB} coincident detections. Assuming that all \ac{NSNS} systems produce short~\acp{GRB} with a jet angle of $20^\circ$, we find that, relative to just an all\nobreakdashes-sky search, adding a search triggered by a \textit{Swift}\nobreakdashes-like satellite increases the total number of \ac{CBC} detections by 2\%, but more importantly doubles the number of \ac{GW}\nobreakdashes--\ac{GRB} associations. A mission such as \textit{Fermi} that has $\sim 6$ times the instantaneous sky coverage of \textit{Swift} would contribute not quite an increase of $\sim 12$\% to the total number of \ac{CBC} detections, because the relatively poor sky localization would permit less of a reduction in the \ac{GW} detection threshold. Although the calculated enhancement of detection rate is dependent on this and other assumptions, we believe it justifies the effort that is being spent on such triggered searches, given the extra scientific value of multimessenger detections.

The externally triggered \ac{GRB} searches to date have attempted to detect a population of sub\nobreakdashes-threshold \acp{GW}. We performed simulations that show that typically there will not be a detectable excess until the rate of direct detections of individual sources is already high; hence it is highly unlikely that an excess\nobreakdashes-population test will provide the first strong evidence for \acp{CBC}.

\begin{acknowledgements}
The authors thank Alan Weinstein and Michal Was for comments on the manuscript, and Neil Gehrels for updating us on \ac{GRB} missions.
\ac{LIGO} was constructed by the California Institute of Technology and
Massachusetts Institute of Technology with funding from the \ac{NSF} and operates under cooperative agreement No. PHY-0107417. L.S. is
supported by the \ac{NSF} through a Graduate Research Fellowship, while A.D. is
supported by \ac{NSF} Grants No. PHY-1067985 and No. PHY-0757937.  C.C.'s work was carried out  at the Jet Propulsion Laboratory, California Institute of Technology, under contract to the National Aeronautics and Space Administration.  C.C. also gratefully acknowledges support from  \ac{NSF} grant PHY1068881.

This paper has LIGO Document No. LIGO-P1200113-v6.
\end{acknowledgements}

\appendix


\section{Prior distribution of \ac{GW} detector network's response to a \ac{CBC} at a given distance and sky location}
\label{app:responses}

In this section we discuss the distribution of a network's response to \acp{GW} from a source at a particular, known sky location and distance, whose orientation is unknown but whose inclination is restricted to be less than a maximum value $\iota_\mathrm{max} = \theta_j$. When studying an individual \ac{GRB}, we could treat this as a prior distribution for the strength of the signal received by a particular detector network. In the event of a nondetection, it would allow us to parameterize the excluded distance by the jet opening angle.

Let the frequency domain \ac{GW} strain received by detector $X$ be $h^X(f)$ and the noise power spectral density of detector $X$ be $S^X(f)$.  Defining the inner product
$$
	(r^X)^2 = \langle h^X, h^X \rangle = 4 \re \int_0^\infty \frac{(h^X)^* (f) h^X (f)}{S^X(f)} \, \mathrm{d}f,
$$
and then the sum over all of the detectors,
$$
	r^2 = \sum_X (r^X)^2,
$$
the coherent detection statistic ${\rho_\mathrm{coh}}^2$ defined by \citet{Harry:2010fr} and the incoherent, coincident detection statistic ${\rho_\mathrm{coinc}}^2$ are noncentrally chi-squared distributed with the noncentrality parameter given by $r^2$.  (The coherent detection statistic has 4 degrees of freedom, whereas the coincident statistic has 2 degrees of freedom times the number of detectors.)  We will first derive summary statistics of $r$: its minimum, maximum, mode, mean, and \ac{rms}.  Then, we will derive the full distribution of $r$ and study its qualitative features.

\citet{Harry:2010fr} introduce $D$, the luminosity distance of the source; $D_0$, an arbitrary fiducial distance; the three angles $(\iota, \psi, \phi_0)$ describing the orientation of the source, being respectively the inclination of the orbital plane to the line of sight, the polarization angle, and the orbital phase at coalescence; and the antenna factors of each detector, $F_+^X$ and $F_\times^X$, which are functions of sky location.  They also define two waveform quadratures, $h_0$ and $h_{\pi/2}$, which are nearly orthogonal such that
$$
	4 \int_0^\infty \frac{|h_0(f)|^2}{S^X(f)} \, \mathrm{d}f \approx 4 \int_0^\infty \frac{|h_{\pi/2}(f)|^2}{S^X(f)} \, \mathrm{d}f = (\sigma^X)^2
$$
and
$$
	4 \re \int_0^\infty \frac{h_0^*(f) h_{\pi/2}(f)}{S^X(f)} \, \mathrm{d}f \approx 0.
$$
They define a further three quantities that combine the antenna factors of a network of detectors,
\begin{equation}\label{eq:abc-def}
	\left.
	\begin{aligned}
		A &= \sum_X (\sigma^X F_+^X)^2 \\
		B &= \sum_X (\sigma^X F_\times^X)^2 \\
		C &= \sum_X (\sigma^X F_+^X)(\sigma^X F_\times^X).
	\end{aligned}
	\right\}
\end{equation}
It is easily---though laboriously---shown that the detector response or noncentrality parameter $r^2$ depends only on the antenna factors, distance, $\iota$, and $\psi$, through
\begin{multline}
	r^2 = \frac{1}{8} \frac{{D_0}^2}{D^2} \bigg[
		(A + B) (x^4 + 6 x^2 + 1) \\
		+ \sqrt{(A - B)^2 + 4 C^2} (1 - x^2)^2 \cos \alpha
	\bigg],
\end{multline}
where $x = \cos \iota$ and $\alpha = 4 \psi - \atan2(2C, A-B)$.  For the purpose of concise parameterization of $r^2$, we will also introduce $x_0 = \cos \iota_\mathrm{max}$, $J^2 = A + B$, and $K^2 = \sqrt{(A - B)^2 + 4 C^2} / (A + B)$.  $J^2$ describes the total sensitivity of the detector network as the weighted sum of squares of all of the antenna factors.  To lend interpretation to $K^2$, we write Eq.~(\ref{eq:abc-def}) as
\begin{align*}
	H &= \left(
		\begin{array}{cccc}
			\sigma^1 F_+^1 & \sigma^2 F_+^2 & \sigma^3 F_+^3 & \cdots \\
			\sigma^1 F_\times^1 & \sigma^2 F_\times^2 & \sigma^3 F_\times^3 & \cdots
		\end{array}
	\right)
	\left(
		\begin{array}{cc}
			\sigma^1 F_+^1 & \sigma^1 F_\times^1 \\
			\sigma^2 F_+^2 & \sigma^2 F_\times^2 \\
			\sigma^3 F_+^3 & \sigma^3 F_\times^3 \\
			\vdots & \vdots \\
		\end{array}
	\right) \\
	&\equiv
	\left(
		\begin{array}{cc}
			A & C \\
			C & B
		\end{array}
	\right).
\end{align*}
Then, $K^2$ can be shown to equal the ratio of the difference of the eigenvalues of $H$ to their sum.  $K^2$ measures the extent to which the network is preferentially sensitive to just one polarization, in a way that is independent with respect to rotations of the detector network's coordinate system.  If the network is equally sensitive to two orthogonal polarizations, then $K^2$ vanishes.  If the network is sensitive to only one polarization, then $K^2$ is unity.

Lastly, we define $u^2$ as the distance\nobreakdashes-independent part of $r^2$:
\begin{align*}
	u^2 &= \left(\frac{{D_0}^2}{D^2}\right)^{-1} r^2 \\
		&= \frac{1}{8} J^2 \left[
		(x^4 + 6 x^2 + 1) + K^2 (1 - x^2)^2 \cos \alpha
	\right].
\end{align*}

Our goal is to study the conditional \ac{pdf}, $p(u \, | \, \theta, \phi, \iota \leq \iota_\mathrm{max})$, of $u$, assuming a fixed sky location and a censored orientation distribution.  Under the assumption that the \ac{GRB} emission is collimated within an angle $\iota_\mathrm{max} = \theta_j$, the burst can only be seen if the Earth is placed inside this cone.  As the probability is the same for random placement anywhere on the surface of this cone, the prior distribution of the inclination angle $\iota$ between the line of sight to Earth and the axis of the outflow is given by
\begin{equation}
\label{eq:prior}
	p(\iota) = \frac{\sin \iota}{1 - \cos \iota_\mathrm{max}}
		\quad \text{where} \quad
		0 \leq \iota \leq \iota_\mathrm{max} \,,
\end{equation}
since an area element on the cone is given by $dA = \sin{\iota}\, d\iota \, d\phi$.
This represents a prior distribution on the direction of the system's orbital axis that is uniform in solid angle, restricted to polar angles $\leq \iota_\mathrm{max}$.

Having parameterized the detector network's response, we proceed to derive the distribution of $r$ for a given sky location and luminosity distance, as well as the minimum, maximum, mean, \ac{rms}, and mode of the distribution. Finally, as an example, we will apply our results to \ac{GRB}~051103.

\subsection{Distribution and summary statistics}

\begin{figure*}[htbp]
\includegraphics[width=\textwidth]{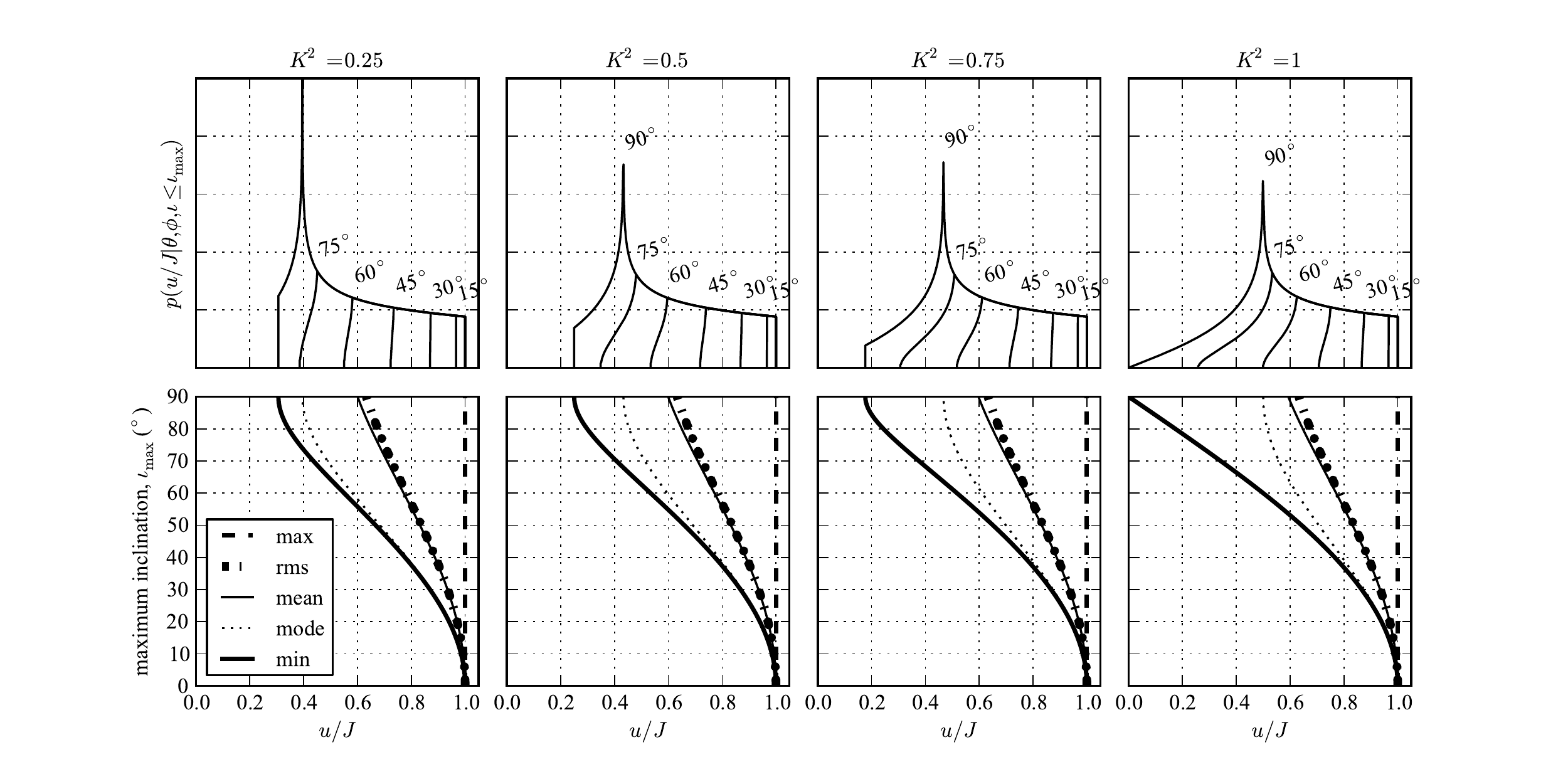}
\caption{\label{fig:pdf}Prior distribution of $(u/J)$, the distance-independent part of the detector response, normalized by the detector's root-sum-squared antenna pattern at a given sky location. The top row shows the probability density function itself, for jet opening angles of $15^\circ$ to $90^\circ$. From left to right, the value of $K^2$ is varied from $0.25$ to $0.1$, smoothly varying from a detector configuration that has similar sensitivity to two polarizations to a configuration that is sensitive to only one polarization. The bottom row of plots shows five summary statistics of the distribution, the minimum, maximum, mean, mode, and \ac{rms}.}
\end{figure*}

The distribution of the detector network's response and its summary statistics, derived below, are plotted in Fig.~\ref{fig:pdf} for a selection of detector networks and maximum inclination angles.

\subsubsection{Minimum and maximum}

The minimum value of $u$ is obtained when the source is at the maximum inclination ($\iota = \iota_\mathrm{max}$ or $x = x_0$) and when $\cos \alpha = -1$:
\begin{equation}\label{eq:umin}
	u_\mathrm{min} = \left(\frac{J}{\sqrt{8}}\right) \sqrt{{x_0}^4 + 6 {x_0}^2 + 1 - K^2 (1 - {x_0}^2)^2}.
\end{equation}
The maximum value is obtained when the source is at the minimum inclination ($\iota = 0$ or $x = 1$):
\begin{equation}\label{eq:umax}
	u_\mathrm{max} = J.
\end{equation}

\subsubsection{Mean}

The mean response is given by
\begin{widetext}
$$
	u_\mathrm{mean} = \int_{x_0}^1 \int_0^{2 \pi} \frac{J}{\sqrt{8}} \frac{\sqrt{(x^4 + 6 x^2 + 1) + K^2 (1 - x^2)^2 \cos \alpha}}{2 \pi (1 - x_0)} \, \mathrm{d}\alpha \, \mathrm{d}x.
$$
It is possible to cast the integral over $\alpha$ into the form of a complete elliptic integral of the second kind,
\begin{equation}\label{eq:umean}
	u_\mathrm{mean} = \frac{J}{\sqrt{2} \pi (1 - x_0)} \int_{x_0}^1 \sqrt{x^4 + 6x^2 + 1 + K^2 (1 - x^2)^2} E \left( \sqrt{\frac{2 K^2 (1 - x^2)^2}{x^4 + 6x^2 + 1 + K^2 (1 - x^2)^2}} \right) \, \mathrm{d}x
\end{equation}
\end{widetext}
where
$$
	E(k) = \int_0^{\frac{\pi}{2}} \sqrt{1 - k^2 \sin^2 \alpha} \, \mathrm{d}\alpha.
$$

\subsubsection{Root-mean-square}

The \ac{rms} response is
\begin{widetext}
$$
	{u_\mathrm{rms}}^2 = \int_{x_0}^1 \int_0^{2 \pi} \frac{J^2}{8} \left[\frac{(x^4 + 6 x^2 + 1) + K^2 (1 - x^2)^2 \cos \alpha}{2 \pi (1 - x_0)}\right] \, \mathrm{d}\alpha \, \mathrm{d}x.
$$
\end{widetext}
The integral over $\alpha$ kills the $\cos \alpha$ term, leaving
$$
	{u_\mathrm{rms}}^2 = \frac{J^2}{8} \int_{x_0}^1 \left[\frac{x^4 + 6 x^2 + 1}{1 - x_0}\right] \, \mathrm{d}x.
$$
The integral over $x$ gives
\begin{equation}\label{eq:urms}
	u_\mathrm{rms} = \left(\frac{J}{2\sqrt{10}}\right) \sqrt{{x_0}^4 + {x_0}^3 + 11{x_0}^2 + 11{x_0} + 16}.
\end{equation}

\subsubsection{Full probability distribution}

We have the prior distribution of $(\iota, \alpha)$,
$$
	p(\iota, \alpha) = \frac{\sin \iota}{2 \pi (1 - \cos \iota_\mathrm{max})}
		\quad \text{where } \alpha \in [0, 2 \pi], \iota \in [0, \iota_\mathrm{max}],
$$
or equivalently,
$$
	p(x, \alpha) = \frac{1}{2 \pi (1 - x_0)}
		\quad \text{where } \alpha \in [0, 2\pi], x \in [x_0, 1].
$$
To compute the conditional \ac{pdf} of $u$, $p(u \, | \, \theta, \phi, \iota \leq \iota_\mathrm{max})$, the first step is to effect a change of variables from $\mathrm{d}x  \, \mathrm{d}\alpha$ to $\mathrm{d}x \, \mathrm{d}u$.  To do this, we write $\mathbf{y} = \mathbf{f} (\mathbf{x})$ where $\mathbf{x} = (x, \alpha)$ and $\mathbf{y} = (x, u)$.  By forming the Jacobian determinant $|\partial \mathbf{f} / \partial \mathbf{x}|$, we find
$$
	\mathrm{d}x \, \mathrm{d}\alpha =
		\left| \begin{array}{cc}
			1 & 0 \\
			\frac{\partial u}{\partial x} & \frac{\partial u}{\partial \alpha}
		\end{array}
		\right|^{-1} \mathrm{d}x \, \mathrm{d}u =
		\left| \frac{\partial u}{\partial \alpha} \right|^{-1}
		\mathrm{d}x \, \mathrm{d}u.
$$
Using the inverse rule for derivatives, $\partial y / \partial x = (\partial x / \partial y)^{-1}$, we solve for $\alpha$ as a function of $u$,
$$
	\alpha = \arccos \frac{8 u^2 / J^2 - (x^4 + 6x^2 + 1)}{K^2 (1 - x^2)^2},
$$
and then differentiate with respect to $u$,
\begin{align*}
	\frac{\partial \alpha}{\partial u} &=
		-\frac{16 u}{J^2 K^2 (1 - x^2)^2}
		\left[1 - \left(\frac{8 u^2 / J^2 -
		(x^4 + 6x^2 + 1)}{K^2 (1 - x^2)^2}\right)^2\right]^{-\frac{1}{2}} \\
	&= -\frac{16 u}{\sqrt{\left[J^2 K^2 (1 - x^2)^2\right]^2
		- \left[8 u^2 - J^2 (x^4 + 6x^2 + 1)\right]^2}}.
\end{align*}
Now we may express the conditional \ac{pdf} that we seek as
$$
	p(u \, | \, \theta, \phi, \iota \leq \iota_\mathrm{max}) = \int_0^{\iota_\mathrm{max}} 2 \left|\frac{\partial \alpha}{\partial u}\right|  p(x, \alpha) \, \mathrm{d}x.
$$
The factor of $2$ accounts for the two distinct values of $\alpha$ that give the same value of $u$.  Altogether,
\begin{widetext}
\begin{equation}
\label{eq:pdf-integral}
	p(u \, | \, \theta, \phi, \iota \leq \iota_\mathrm{max}) =
		\frac{16}{\pi J (1 - x_0)} \int_{x_1}^{x_2} \frac{u/J}{\sqrt{%
			\left[K^2 (1 - x^2)^2\right]^2
			- \left[8 (u/J)^2 - (x^4 + 6x^2 + 1)\right]^2}}
		\, \mathrm{d}x.
\end{equation}
\end{widetext}
We have to be a little careful about the limits of integration; the quantity in the radical must remain positive.  It has a zero at
$$
	x^2 = \frac{-3 + K^2 + \sqrt{8[1 - K^2 + u^2 (1 + K^2) / J^2]}}{1 + K^2}.
$$
The lower limit should be
\begin{equation}
\label{eq:x1}
	x_1 = \sqrt{\max \left(
		\frac{-3 + K^2 + \sqrt{8[1 - K^2 + \left(\frac{u}{J}\right)^2 (1 + K^2)]}}{1 + K^2},
		{x_0}^2\right)}.
\end{equation}
The upper limit should be
\begin{equation}
\label{eq:x2}
	x_2 = \begin{cases}
		\sqrt{\frac{-3 - K^2 + \sqrt{8[1 + K^2 + u^2 (1 - K^2) / J^2]}}{1 - K^2}} & \text{if } K \neq 1 \\
		u / J & \text{if } K = 1
	\end{cases}
\end{equation}
Eq.~(\ref{eq:pdf-integral}) may be evaluated numerically using, for example, Simpson's rule.

\subsubsection{Mode}

The mode of the distribution occurs when the lower limit of integration ceases to clip against the minimum value, where
$$
	x_0 = \sqrt{\frac{-3 + K^2 + \sqrt{8[1 - K^2 + u^2 (1 + K^2) / J^2]}}{1 + K^2}}.
$$
This occurs at
\begin{equation}\label{eq:umode}
	u_\mathrm{mode} = \left(\frac{J}{\sqrt{8}}\right) \sqrt{{x_0}^4 + 6 {x_0}^2 + 1 + K^2 (1 - {x_0}^2)^2}.
\end{equation}

\subsection{Special case: Unrestricted inclination}

If the inclination is unrestricted, $\iota_\mathrm{max} = \pi/2$ or $x_0 = 0$, then the above results simplify to
\begin{align*}
	u_\mathrm{mode} &= J \sqrt{1 + K^2} / \sqrt{8} \\
	u_\mathrm{min}  &= J \sqrt{1 - K^2} / \sqrt{8} \\
	u_\mathrm{max}  &= J \\
	u_\mathrm{rms}  &= \sqrt{2} J / \sqrt{5}.
\end{align*}
Neither $u_\mathrm{mean}$ nor the full \ac{pdf} $p(u \, | \, \theta, \phi, \iota \leq \iota_\mathrm{max})$ simplify much for the unrestricted inclination case.

\subsection{Special case: One detector}

When the detector network consists of only one detector, $C^2 = A B$, $K = 1$, and $J^2 = \sigma^2 F^2$, where $F^2 = {F_+}^2 + {F_\times}^2$.  With these substitutions,
\begin{widetext}
\begin{align*}
	u_\mathrm{mode} &= J (1 + {x_0}^2) / 2 \\
	u_\mathrm{min}  &= J x_0 \\
	u_\mathrm{max}  &= J \\
	u_\mathrm{rms}  &= \left(\frac{J}{2\sqrt{10}}\right) \sqrt{{x_0}^4 + {x_0}^3 + 11{x_0}^2 + 11{x_0} + 16} \\
	u_\mathrm{mean} &= \frac{J}{\pi (1 - x_0)} \int_{x_0}^1 (1 + x^2) E \left( \frac{1 - x^2}{1 + x^2} \right) \, \mathrm{d}x \\
	p(u \, | \, \theta, \phi, \iota \leq \iota_\mathrm{max}) &=
		\frac{16}{\pi J (1 - x_0)} \int_{x_1}^{x_2} \frac{u/J}{\sqrt{%
			\left[(1 - x^2)^2\right]^2
			- \left[8 u^2/J^2 - (x^4 + 6x^2 + 1)\right]^2}}
		\, \mathrm{d}x \\
	x_1 &= \sqrt{\max \left(
		2 u / J - 1,
		{x_0}^2\right)} \\
	x_2 &= u / J.
\end{align*}
\end{widetext}

\subsection{Case study: \ac{GRB}~051103}

\ac{GRB}~051103 was an exceptionally short, hard, and bright burst detected by \textit{HETE}, \textit{Suzaku}, and \textit{Swift}, and localized by \ac{IPN} to an area consistent with the outer disc of M81~\citep{MNR:MNR16118}. Owing to its brightness and hardness, a giant flare from an extragalactic \ac{SGR} was a plausible progenitor. The \ac{H2} and \ac{L1} were operating at the time, so a targeted search of the \ac{GW} data was undertaken. No candidate was detected, but the nondetection excluded a \ac{CBC} event in M81 as the progenitor \citep{2012ApJ...755....2A}. Under the assumption that a \ac{CBC} progenitor would have produced a collimated jet along the axis of strongest gravitational wave emission, \citet{2012ApJ...755....2A} placed 90\% confidence lower limits on the distance of a \ac{CBC} progenitor as function of jet angle. A collimated \ac{GRB} in M81 was firmly excluded.

As an example, we apply our distribution of detector response to the problem of estimating the exclusion distance as a function of jet opening angle for \ac{GRB}~051103. If we knew the \ac{GW} search's detection efficiency for strictly face\nobreakdashes-on sources, then using our distribution for $(u/J)$ we could directly calculate the excluded distance for any jet opening angle and any confidence level. \citet{2012ApJ...755....2A} did not publish that detection efficiency, but we can do a qualitatively similar calculation by extrapolating from their 90\% exclusion distance for $\theta_j = \iota_\mathrm{max} = 10^\circ$, attempting to reproduce their exclusion distance at other jet opening angles.

\ac{GRB}~051103 occurred at 3 November 2011 09:25:42 UTC. The \ac{H2} and \ac{L1} horizon distances (distance at which an optimally oriented face-on \ac{CBC} would register an amplitude $r=8$) at this time for both a 1.4\nobreakdashes--1.4~$M_\odot$ \ac{NSNS} event and a 1.4\nobreakdashes--10~$M_\odot$ \ac{NSBH} event are given in Table~\ref{tab:antenna}, along with the antenna factors at this time in the direction of M81. For a \ac{NSNS} signal, this network has $K^2 = 0.9601$, and for a \ac{NSBH} signal, $K^2 = 0.9602$. \ac{H2} and \ac{L1} had almost the same sensitivity up to a frequency-independent factor of $\approx$2, so it is not surprising that the value of $K^2$ is almost the same for both the \ac{NSNS} and \ac{NSBH} signal models.

\begin{figure}[htbp]
\includegraphics{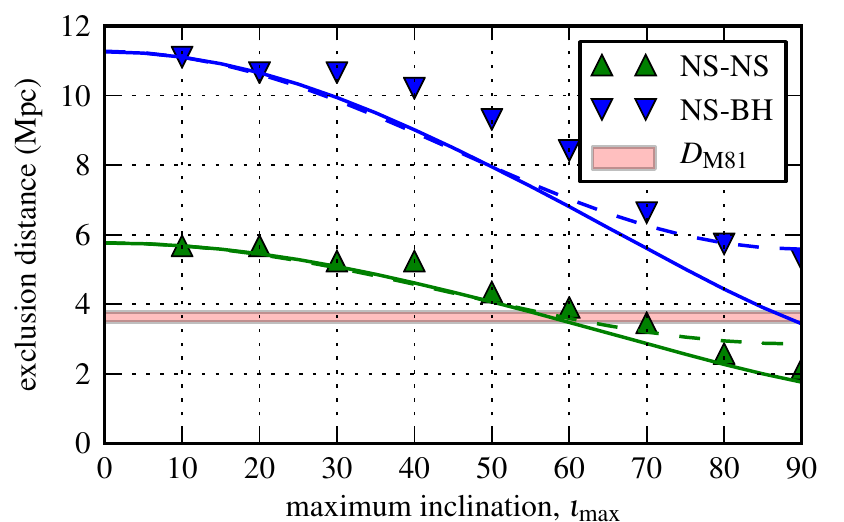}
\caption{\label{fig:exclusion}Exclusion distance as a function of $\iota_\mathrm{max} = \theta_j$. The filled triangles are the 90\%\nobreakdashes-confidence exclusion distances from Fig.~3 of \citet{2012ApJ...755....2A}. The up arrows (green in the online version) represent the \ac{NSNS} signal model, and the down arrows (blue in the online version) represent the \ac{NSBH} signal model. The solid curves show the 10\% value of the inverse \ac{CDF} of $(u/J)$, scaled to match the exclusion distance at $\iota_\mathrm{max}=10^\circ$. The dashed curves show the mode of the distribution of $(u/J)$, also scaled to match the exclusion distance at $\iota_\mathrm{max}=10^\circ$. The pink band marks the distance to M81, $3.63\pm0.14$ Mpc.}
\end{figure}

In Fig.~\ref{fig:exclusion}, we plot the 90\% exclusion distance as a function of $\theta_jet$ from Fig.~3 of \citet{2012ApJ...755....2A}. We have superimposed the mode of the detector response distribution, Eq.~(\ref{eq:umode}), scaled to match the published exclusion distance at $\theta_j=10^\circ$, as a dashed line. The value of $(u/J)$ at which the \ac{CDF} is equal to $(1 - 0.9)$ is shown as a solid line. The inverse \ac{CDF} agrees well with the \ac{NSNS} exclusion distance, but the mode agrees much better with the \ac{NSBH} exclusion distance than the inverse \ac{CDF}. Exact agreement is not expected with either: as we have pointed out, a proper application to calculating exclusion distances would require knowledge of both the prior distribution of $(u/J)$ and the sensitivity of the \ac{GW} search to face-on sources as a function of signal amplitude. Furthermore, the analysis of \citet{2012ApJ...755....2A} includes a Monte Carlo integration over a range of masses, whereas our analysis fixes canonical choices of the masses.

\begin{table}
\caption{\label{tab:antenna}Antenna factors and horizon distances for \ac{H2} and \ac{L1} detectors at time of \ac{GRB}~051103.}
\begin{ruledtabular}
\begin{tabular}{rdddd}
Detector &
\multicolumn{1}{c}{$F_+$} &
\multicolumn{1}{c}{$F_\times$} &
\multicolumn{1}{c}{$DH_\mathrm{NSNS}$} &
\multicolumn{1}{c}{$DH_\mathrm{NSBH}$} \\
\colrule
\ac{H2} & -0.152 & -0.706 & 8.3 & 17.1 \\
\ac{L1} & 0.348 & 0.550 & 18.1 & 37.4 \\
\end{tabular}
\end{ruledtabular}
\end{table}

\section{Enhanced \ac{GW} amplitude of \ac{GRB}-triggered sources}
\label{app:GWamp}

In this paper, we assume that the emitted gamma rays are beamed within an
angle $\theta_j$ of the normal to the orbital plane and thus the
binary inclination $\iota$ must be less than $\theta_j$. Where
$\theta_j$ is small, we approximate the \ac{GW} amplitude for
\ac{GRB}-triggered sources to be on average $\sim 1.51$ times the isotropic
detection-averaged amplitude for all \acp{CBC}, which is the instantaneous
$\iota=0$ value.

Using intermediate results from Appendix~\ref{app:responses}, the azimuthally
averaged detector response to a binary whose orbital plane inclined at angle
$\iota$ relative to the observer's line of sight is proportional to
\begin{equation}
\left(\frac{u}{J}\right)^2 = \frac{1}{8} \left(x^4 + 6 x^2 + 1 \right)\,,
\end{equation}
where $x = \cos\iota$ and $J^2 = F^2_+ + F^2_\times$.  The distance to which a
\ac{GW} source is detectable scales as $u/J$, so the number of detectable
sources scales as $(u/J)^3$.  Thus the detection-averaged amplitude of all the
sources that we observe within half-angle $\theta_j$ is
\begin{equation}
\label{Aenhance}
\bar A(x_0) = \left[\int_{x_0}^1 \left(\frac{u}{J}\right)^3 p(x) dx\right]^{1/3}\,,
\end{equation}
where $x_0 = \cos \theta_j$ and $p(x) = (1 - x_0)^{-1}$.

Using Eq.~\ref{Aenhance}, one easily shows that as long as the beam half-angle
is $\lesssim 25^{\circ}$, then average amplitude enhancement relative to an
unrestricted distribution is at most a $\sim 5\%$ overestimate.

\bibliographystyle{apsrev}
\bibliography{../../bibtex/extra_aps_macros,GRBoverview,../../bibtex/iulpapers}

\end{document}